\documentclass[twocolumn,prl,aps,showpacs]{revtex4}
\usepackage{graphicx,amsmath,amssymb,mathptmx}
\newcommand{\abbrev}[1]{\mbox{\small{#1}}}
\newcommand{\ep}{\epsilon}
\newcommand{\eqn}[1]{Eq.\,(\ref{#1})}

\newcommand{\reference}[1]{Ref.\,\cite{#1}}

\renewcommand{\L}{\left(}
\newcommand{\R}{\right)}

\newcommand{\api}{\frac{\alpha_s}{\pi}}
\newcommand{\nlo}{\abbrev{NLO}}
\newcommand{\nnlo}{\abbrev{NNLO}}
\newcommand{\qcd}{\abbrev{QCD}}
\newcommand{\cm}{\rm c.m.}
\newcommand{\lab}{\rm lab}
\newcommand{\ie}{{\it i.e.\/}}
\newcommand{\qb}{{\overline q}}

\begin{document}

\begin{titlepage}

\hspace*{\fill}\parbox[t]{2.6cm}{
BNL-HET-04/3\\
NSF-KITP-04-34\\
11 March 2004
}

\bibliographystyle{apsrev}
\preprint{BNL-HET-04/3}
\preprint{NSF-KITP-04-34}
\eprint{hep-ph/0403128}

\title{Subtraction Terms for Hadronic Production Processes at
  Next-to-Next-to-Leading Order}

\author{William~B.~Kilgore}

\affiliation{Physics Department, Brookhaven National Laboratory,
  Upton, New York
  11973, U.S.A.\\
  {\tt [kilgore@bnl.gov]} }

\begin{abstract}
  I describe a subtraction scheme for the next-to-next-to-leading
  order calculation of single inclusive production at hadron
  colliders.  Such processes include Drell-Yan, $W^\pm$, $Z$ and Higgs
  Boson production.  The key to such a calculation is a treatment of
  initial state radiation which preserves the production
  characteristics, such as the rapidity distribution, of the process
  involved.  The method builds upon the Dipole Formalism and, with
  proper modifications, could be applied to deep inelastic scattering
  and $e^+e^-$ annihilation to hadrons.
\end{abstract}

\maketitle
\end{titlepage}


\section{Introduction}\label{sec::Intro}
Next-to-next-to-leading order (\nnlo) calculations combine three
contributions: Second order virtual corrections, first order virtual
corrections to single real emission and double real emission.  For
special processes, like inclusive single particle production or
inclusive deep inelastic scattering, fully inclusive \nnlo\
calculations can be performed by analytically integrating over the
total phase space~\cite{Hamberg:1991np, vanNeerven:1991nn,
Harlander:2002wh, Anastasiou:2002yz, Harlander:2002vv,
Anastasiou:2002wq, Ravindran:2003um, Harlander:2003ai,
Ravindran:2003gi}.  In doing so, however, much of the exclusive
information from the real emission processes is lost.  In addition, it
is difficult to impose geometric and kinematic cuts on the phase space
that would correspond to a realistic experimental environment,
although important progress within the framework of totally inclusive
calculations has recently been made on this
subject~\cite{Anastasiou:2003yy, Anastasiou:2003ds}.  Still, one would
like to be able to perform a numerical calculation, valid to \nnlo,
which would retain the exclusive information of hard real emission and
allow one to impose arbitrarily complicated cuts on the data.

The difficulty in constructing such a program is that each of the
three contributions to the \nnlo\ cross section is infrared divergent.
It is only the sum of the three that yields a meaningful physical
result.  Combining the three terms is made difficult by the fact that
each term involves a different number of final state particles and
must therefore be integrated over a different phase space.  Still, it
should be possible to reorganize the infrared singularities, by adding
and subtracting local counter-terms, so that one can perform and
combine three {\it finite} calculations.  Indeed, at next-to-leading
order (\nlo), the solution is well understood and a variety of methods
have been successfully implemented.

A great deal of work has been devoted to the development of a general
algorithm for constructing the local counter-terms at
\nnlo~\cite{Kosower:2002su, Kosower:2003cz, Kosower:2003bh,
Weinzierl:2003fx, Weinzierl:2003ra, Weinzierl:2003tk}, but as yet
there is no concrete example of a successful algorithm.  Indeed the
first successful ``semi-inclusive'' \nnlo\ calculation was
accomplished very recently~\cite{Anastasiou:2004qd}, and did not use
the analytic cancellation of infrared singularities afforded by a
general subtraction method.  Instead, it used the strategy of sector
decomposition~\cite{Binoth:2000ps, Binoth:2003ak,
Gehrmann-DeRidder:2003bm, Anastasiou:2003gr, Binoth:2004jv} to
numerically cancel the singular contributions.

In this letter, I will describe an analytic subtraction scheme for the
\nnlo\ calculation of inclusive single particle production which
builds upon the framework of the dipole formalism~\cite{Catani:1996jh,
Catani:1997vz}.  In doing so, I do not solve the hardest problems
associated with constructing a general purpose \nnlo\ algorithm.
Instead, I make use of the relative simplicity of the amplitudes
involved and the control afforded by my ability to compute total
integrals in each of the phase spaces.


\section{The Basic Framework at \nlo}\label{sec::Framework}
I first review the basic framework for \nlo\ calculations and
build upon that to construct a solution at \nnlo.  An \nlo\ calculation
for an $n$ parton cross section can be expressed as
\begin{equation}
\sigma^{\nlo} = \int_{n+1} d\sigma^{(0)}_{n+1} + \int_{n}
 d\sigma^{(1)}_{n},
\label{eqn:nlobasic}
\end{equation}
where the subscripts refer to the number of final state particles for
a particular term and the superscripts refer to the order in the
$\alpha_s$ expansion to which the term is calculated.  Because of the
different numbers of particle in their phase spaces, the two
calculations must be performed separately.  The fact that both
contributions are infrared divergent presents a non-trivial challenge
to numerical calculations.  A solution to this problem involves the
construction of local counter-terms to the real emission cross section
which permit a factorization of the phase space.  The construction of
such local counter-terms is made possible by the infrared
factorization properties of \qcd\ matrix elements.  Introducing such
terms, the \nlo\ calculation now becomes
\begin{equation}
\begin{split}
\sigma^{\nlo}
     &= \int_{n+1} \L d\sigma^{(0)}_{n+1} - d\alpha^{(0)}_{n+1}\R\\
     &+ \int_{n} d\sigma^{(1)}_{n} + \int_{n+1}
             \kern-10pt d\alpha^{(0)}_{n+1}\,,
\end{split}
\label{eqn:nlosubtract}
\end{equation}
where $d\alpha^{(0)}_{n+1}$ is the local counter-term.
$d\alpha^{(0)}_{n+1}$ is constructed such that it factorizes into the
product of $d\sigma^{(0)}_{n}$, the $n$ parton Born term, and an
infrared singular term that can be completely integrated out of a
factorization of $(n+1)$-body phase space into $n$-body phase space
times $1$-body phase space.  Thus, the terms on the first line in
\eqn{eqn:nlosubtract} are evaluated in the $(n+1)$-body phase space,
while those on the second line are evaluated in the $n$-body phase
space.

When the radiative emission is hard, or rather, resolved,
$d\alpha^{(0)}_{n+1}$ and $d\sigma^{(0)}_{n+1}$ map to different
points in phase space and may be quite different in numerical value.
When the emission is unresolved, however, the two terms map to the
same point in phase space and their infrared singular terms cancel
numerically.  In the limit, the cancellation is complete and such
terms do not contribute to the integral.



\section{On to \nnlo}\label{sec::NNLO}
In analogous fashion, an \nnlo\ calculation is expressed as
\begin{equation}
\sigma^{\nnlo} = \int_{n+2} d\sigma^{(0)}_{n+2} + \int_{n+1}
d\sigma^{(1)}_{n+1} + \int_{n} d\sigma^{(2)}_{n}.
\label{eqn:nnlobasic}
\end{equation}
Again, each of the integrals on the right hand side is infrared
divergent and again, one proceeds by constructing local counter-terms,
but now the structure of the terms will be rather more complicated.
\begin{equation}
\begin{split}
\sigma&\vphantom{\sigma}^{\nnlo} =
   \int_{n+2} \L d\sigma^{(0)}_{n+2} - d\alpha^{(0)}_{n+2}
       + d\beta^{(0)}_{n+2} - d\gamma^{(0)}_{n+2}\R\\
     & + \int_{n+1}\L d\sigma^{(1)}_{n+1} - d\alpha^{(1)}_{n+1}\R
       + \int_{n+2}\L d\alpha^{(0)}_{n+2} - d\beta^{(0)}_{n+2}\R\\
     & + \int_{n} d\sigma^{(2)}_{n} + \int_{n+1}d\alpha^{(1)}_{n+1}
       + \int_{n+2} d\gamma^{(0)}_{n+2}\,,
\end{split}
\label{eqn:nnlosubtract}
\end{equation}
where $d\alpha^{(0)}_{n+2}$ is the tree-level single real emission
counter-term to $d\sigma^{(0)}_{n+2}$ like that found at \nlo,
$d\beta^{(0)}_{n+2}$ is a local counter-term to $d\alpha^{(0)}_{n+2}$
which cancels the singularities due to a subsequent second real
emission, $d\gamma^{(0)}_{n+2}$ is the tree-level double real emission
counter-term to $d\sigma^{(0)}_{n+2}$, and $d\alpha^{(1)}_{n+1}$ is
the one-loop single real emission counter-term to
$d\sigma^{(1)}_{n+1}$.  As I have written \eqn{eqn:nnlosubtract}, the
terms on the first line are all computed in the $(n+2)$-body phase
space, the terms on the second line in $(n+1)$-body phase space (after
integrating out a single emission from $d\alpha^{(0)}_{n+2}$ and
$d\beta^{(0)}_{n+2}$) and the terms on the third line are computed in
$n$-body phase after integrating out the single and double emissions
from $d\alpha^{(1)}_{n+1}$ and $d\gamma^{(0)}_{n+2}$ respectively.

It might seem that $d\beta^{(0)}_{n+2}$ could be constructed by
iterating the procedure used to produce $d\alpha^{(0)}_{n+2}$.  This
is unlikely to work properly since only the infrared structure of each
factorization is universal.  The overlap of infrared singularities
from one factorization with the finite remainder of the other is
likely to generate spurious infrared divergences.  Moreover it is not
necessary.  One does not need a factorized approximation to
$d\alpha^{(0)}_{n+2}$ since one does not need to integrate out the
second emission to get down to the $n$-body phase space.  Instead, one
merely needs to know the locus in phase space that
$d\alpha^{(0)}_{n+2}$ will map to under sequential emission.  Thus,
$d\beta^{(0)}_{n+2}$ should be made numerically identical to
$d\alpha^{(0)}_{n+2}$ but should be evaluated at the point in phase
space corresponding to the second emission.

The $d\gamma^{(0)}_{n+2}$ and $d\alpha^{(1)}_{n+1}$ counter-terms must
approximate the soft and collinear limits of one-loop single real
emission~\cite{Bern:1999ry, Kosower:1999rx, Catani:2000pi} and double
real emission~\cite{Berends:1989zn, Catani:1992XX, Campbell:1998hg,
Catani:1998nv, Catani:1999ss}.  Recently, Weinzierl has reported a
result for $d\alpha^{(1)}_{n+1}$ appropriate for one-loop final-state
emission, which is the only case needed for the computation of
$e^+e^-\to$ jets~\cite{Weinzierl:2003ra}.

It would be possible to reformulate Weinzierl's subtraction terms for
case of initial state radiation required for the single particle
production processes considered here, but that is not the solution
that I propose.  Instead, I use the same strategy in constructing
$d\gamma^{(0)}_{n+2}$ and $d\alpha^{(1)}_{n+1}$ as I used for
constructing $d\beta^{(0)}_{n+2}$: rather than constructing
approximations to the matrix element at some point in phase space, I
use the exact matrix elements but evaluate them as if they were at
different point in phase space.  I can do this because I have complete
analytic control of the total integrals of these matrix elements over
phase space.

With this strategy, it is clear that if I compute the total rate,
placing no kinematic or geometric cuts on the configuration and not
binning any distributions, the first two lines in
\eqn{eqn:nnlosubtract} vanish identically, while the third line gives
the known result for the inclusive cross section.


\section{Rapidity Distributions}\label{sec::Rapidity}
The simple framework described above is sufficient for describing the
total rate, but does not include all of the information available in
inclusive production.  In addition to the total rate, one can also
observe the rapidity distributions of the vector or Higgs
boson~\cite{Anastasiou:2003yy, Anastasiou:2003ds}.  In order to
reproduce rapidity distributions in this calculation, one needs
rapidity information in the subtraction term.  The subtraction term in
the full $(n+2)$-body phase space certainly contains this information.
What is needed is a means of capturing this information in the
$n$-body phase space.  The way to do so is to understand the structure
of the subtraction term in the $n$-body phase space: in final state
emission, the subtraction term in the $n$-body phase space is
essentially a number; in initial state emission, it is a convolution
of the emission contributions with the partonic cross section and can
be expressed in a form reminiscent of mass factorization,
\begin{equation}
\begin{split}
  \sigma^{(m)}_{n+2-m} &= 
  \sum_{k=0}^{m}\sum_{j=0}^{2-k-m}\sigma^{(k)}_{n}\otimes\,
  \tilde\Gamma^{(j)}_{1}\otimes\,\tilde\Gamma^{(2-j-k-m)}_{2}\,,
\end{split}
\label{eqn:massfact}
\end{equation}
where the superscripts refer to the order in the expansion
in $\alpha_s$ of each term.

Unlike mass factorization, which relates the (infrared singular)
contributions determined by sum of squared matrix elements integrated
over phase space to a convolution of finite partonic cross section
with the mass factorization counter-terms associated with the
Altarelli-Parisi splitting functions, \eqn{eqn:massfact} seeks to
relate individual components, say $q\qb\to Vgg$, to a convolution of
the virtual cross section, evaluated to the appropriate order, with
real emission counter-terms.  The ``appropriate order'' is determined
by the factorization properties of the \qcd\ matrix elements.
Specifically, double real radiation terms map onto a convolution of
the virtual term at Born level with either two single emission terms
or one double emission term, while one-loop single real radiation
terms map onto the sum of the one-loop virtual term convolved with one
first-order single emission term and the Born level virtual term
convolved with one second-order single emission term.

To be more explicit, I will consider vector boson production and
define the mass of the vector boson to be $M_V$ and the center-of-mass
(\cm) energy squared of the production process to be $\hat{s}$.  The
ratio of the vector boson mass squared to $\hat{s}$ is defined to be
$z\equiv {M_V^2}/\hat{s}$, the momentum fractions of the incoming
partons are defined to be $x_1$ and $x_2$ ($\hat{s} = x_1\,x_2\,s$),
and the fraction of those momentum fractions that go into vector boson
production are defined to be $w_1$ and $w_2$ ($z = w_1\,w_2$).  In
terms of these parameters,
\begin{equation}
\begin{split}
d\hat\sigma_{n}(z) &= \delta(1-z)
   \L a_0 + \api a_1 + \L\api\R^2 a_2\R\,,\\
\tilde\Gamma_i(w_i) &= \delta(1-w_i)
   + \api\tilde\Gamma_i^{(1)}(w_i) 
   + \L\api\R^2 \tilde\Gamma_i^{(2)}(w_i)\,.
\end{split}
\label{eqn:apiseries}
\end{equation}

In the virtual terms, all of the energy goes into vector boson
production and the vector boson rapidity is identical to the
rapidity of the \lab\ system.
\begin{equation}
y_{n} = \frac{1}{2}\ln\frac{x_1}{x_2}\,. 
\label{eqn:virtrap}
\end{equation}

In the case of single real radiation, some of the \cm\ energy goes
into vector boson production, but some goes into a single collinear
emission.  The convolution consists of terms for which the emission is
in the direction of parton $1$ and terms for which it is in the
direction of parton $2$.  Given the energy and boost of the \cm\
system, the mass of the vector boson and the direction of the
radiative emission, it is a simple exercise to determine the rapidity
of the vector boson:
\begin{equation}
y_{n+1} = \frac{1}{2}\ln\frac{x_1}{x_2} \pm \frac{1}{2}\ln z\,.
\label{eqn:singlerap}
\end{equation}
In an actual calculation, one generates vector bosons at a given
rapidity and then convolves the parton distributions over the parton
momentum fractions, using $\tilde\Gamma$ to properly weight the
contributions above threshold.

Things are more complicated when one must consider double emission.
In the triple collinear limit, where both emissions come from the same
incoming leg, the kinematics is unchanged from the single emission
case and one can determine the rapidity of the vector boson from the
direction of the emission.  The case of overlapping emission, where
each incoming parton contributes to the radiation is more complicated.
In order to determine the rapidity of the vector boson, one needs to
know what fraction of the radiation came from each side.  Thus, for
given parton momentum fractions $x_1$, $x_2$, rather than generating
vector bosons at distinct rapidities, as in \eqn{eqn:singlerap}, one
generates them over a continuum of rapidities:
\begin{equation}
\begin{split}
y_{n+2} &= \frac{1}{2}\ln\frac{x_1\,w_1}{x_2,w_2}\,,\\
\frac{1}{2}\ln\frac{x_1}{x_2} + \frac{1}{2}\ln z\,& \le\, y_{n+2}\, \le\,
    \frac{1}{2}\ln\frac{x_1}{x_2} - \frac{1}{2}\ln z\,.\\
\end{split}
\label{eqn:doublerap}
\end{equation}
Again, one generates vector bosons of a given rapidity and then
convolves the parton distributions over the momentum fractions using
$\tilde\Gamma$ to properly weight the contributions above threshold,
but now one needs to know the appropriate $\tilde\Gamma_i$ on each
side.  The total integral over double emission phase space does not
give $\tilde\Gamma_i$, however, it gives the second order contribution
to the convolution
\begin{equation}
\tilde\Gamma^{(2)}(z) = \tilde\Gamma^{(2)}_1(z)
      +\,\tilde\Gamma^{(2)}_2(z)
      +\,\tilde\Gamma^{(1)}_1(w_1)\otimes\,\tilde\Gamma^{(1)}_2(w_2)\,.
\label{eqn:dirad}
\end{equation}

The terms $\tilde\Gamma^{(2)}_i$ correspond to the triple collinear
limits of the incoming partons.  Subtracting these terms off, one is
left with just the overlapping emission term, each component of which
can be expanded as a Laurent series in $\ep$,
\begin{equation}
\begin{split}
\tilde\Gamma^{(2)}_{\rm overlap} &=
  \,\tilde\Gamma^{(1)}_1\otimes\,\tilde\Gamma^{(1)}_2\,,\\
\tilde\Gamma^{(2)}_{\rm overlap} &= \frac{g_{(-4)}}{\ep^4}
  + \frac{g_{(-3)}}{\ep^3} + \frac{g_{(-2)}}{\ep^2}
  + \frac{g_{(-1)}}{\ep} + g_{(0)} + \ldots\,,\\
\tilde\Gamma^{(1)}_i &= \frac{a_{i,(-2)}}{\ep^2} + \frac{a_{i,(-1)}}{\ep}
  + a_{i,(0)} + \ep\,a_{i,(1)} + \ep^2\,a_{i,(2)} + \ldots\,.
\end{split}
\label{eqn:overlap}
\end{equation}
Since the infrared structure of single emission is universal, the most
singular terms in \eqn{eqn:overlap} are fixed and the less singular
terms can be solved for, term by term.  When single emission is
considered in isolation, the nonsingular terms, $a_{i,n\ge0}$, are
non-universal.  In the context of overlapping divergences from double
emission, however, these terms can be given meaningful definitions.
\begin{equation}
\begin{split}
g_{(-4)} &= a_{1,(-2)}\otimes\,a_{2,(-2)}\,,\\
g_{(-3)} &= a_{1,(-1)}\otimes\,a_{2,(-2)}
    + a_{1,(-2)}\otimes\,a_{2,(-1)}\,,\\
g_{(-2)} &= a_{1,(0)}\otimes\,a_{2,(-2)}
    + a_{1,(-2)}\otimes\,a_{2,(0)}\\
         &\kern15pt + a_{1,(-1)}\otimes\,a_{2,(-1)}\,,\\
g_{(-1)} &= a_{1,(1)}\otimes\,a_{2,(-2)}
    + a_{1,(-2)}\otimes\,a_{2,(1)}\\
         &\kern15pt + a_{1,(0)}\otimes\,a_{2,(-1)}
    + a_{1,(-1)}\otimes\,a_{2,(0)}\,,\\
g_{(0)} &= a_{1,(2)}\otimes\,a_{2,(-2)}
    + a_{1,(-2)}\otimes\,a_{2,(2)}\\
         &\kern15pt + a_{1,(1)}\otimes\,a_{2,(-1)}
    + a_{1,(-1)}\otimes\,a_{2,(-1)}\\
         &\kern15pt + a_{1,(0)}\otimes\,a_{2,(0)}\,,\\
\end{split}
\end{equation}
The singular terms must cancel against other infrared contributions,
so all components involved in those terms can be determined
unambiguously.  While the soft contributions (corresponding to delta
functions in the parton fractions) cannot be uniquely distributed
among the separate terms, their distribution does not affect the
magnitude of the total subtraction nor its location in phase space.
The only room for ambiguity would come in the terms that are new to
$g_{(0)}$, \ie\ $a_{1,(2)}$ and $a_{2,(2)}$.  When the splittings are
symmetric, say for $q\qb\to Vgg$ each splitting would be
$q\to\check{q}g$ (where $\check{a}$ identifies the incoming parton),
the solution is symmetric and there is no ambiguity.  When the
splittings are asymmetric, say for $qg\to Vqg$ where one splitting is
$q\to\check{q}g$ and the other is $q\to\check{g}q$ (note the order),
it may seem that one cannot separate the new terms.  However, the
prior knowledge of the more singular terms with which they are
convolved should allow for this extraction.  Certainly the dominant
terms can be determined unambiguously. It is also possible to obtain
rapidity distributions at \nnlo\ within the framework of inclusive
calculations and this has already been done for Drell-Yan
production~\cite{Anastasiou:2003ds}.  The information from such a
calculation could be used to resolve any lingering ambiguities.

Implicit to the preceding discussion is that the $d\gamma^{(0)}_{n+2}$
and $d\alpha^{(1)}_{n+1}$ in the $(n+2)$ and $(n+1)$-body phase spaces
must be evaluated at points in phase space that correspond to those
described in $n$-body phase space.  Those points correspond to
conserving the incoming parton momentum fractions of the true process,
but mapping all final state momenta onto the beam axes.  In the event
of unresolved emission, this mapping ensures a proper subtraction of
the singularities.


\section{Conclusions}
In this letter, I have described the explicit construction of the
subtraction terms needed for an \nnlo\ parton-level monte carlo
calculation of single inclusive production (Higgs bosons, vector
bosons, Drell-Yan) within the dipole formalism.  For these simple
processes, one has complete analytic control over the total integrals
of the matrix elements.  This control allows one to use the matrix
elements themselves as the subtraction terms.  In order to capture the
full information available from inclusive calculations, one needs to
retain information on both the total rate and the rapidity
distribution of production.  I have outlined a method for doing so.

\paragraph*{Acknowledgments:}
I would like to thank Michelangelo Mangano for emphasizing the
importance of this project.  I would also like to thank the Kavli
Institute for Theoretical Physics, where part of this work was done,
for their kind hospitality.  This research was supported in part by
the National Science Foundation under Grant No.~PHY99-07949 and in
part by the U.~S.~Department of Energy under Contract
No.~DE-AC02-98CH10886.

\paragraph*{Note:}
As this paper was being completed, I became aware of
\reference{Gehrmann-DeRidder:2004tv}, in which essentially the same
solution is proposed for the final state emission process, $e^+e^-\to
2$ jets, and is worked out in great detail.



\end{document}